\documentclass[conference]{IEEEtran}

\usepackage{amsfonts}
\usepackage{amssymb}
\usepackage{stfloats}
\usepackage{cite}
\usepackage{graphicx}
\usepackage{psfrag}
\usepackage{subfigure}
\usepackage{amsmath}
\usepackage{array}
\usepackage{algorithm}
\usepackage{algpseudocode}
\usepackage{setspace}
\usepackage[english]{babel}
\usepackage{blindtext}
\usepackage{url}
\usepackage{epstopdf}
\usepackage{verbatim}
\usepackage{color}
\usepackage{amsmath}
\usepackage{bm}
\usepackage{multirow}
\usepackage{booktabs}
\usepackage{amsthm}

\newtheorem{Rem}{Remark}
\newcommand{\RNum}[1]{\uppercase\expandafter{\romannumeral #1\relax}}

\title{Dynamic Carrier and Power Amplifier Mapping for Energy Efficient Multi-Carrier Wireless Communications}

\author{Shunqing Zhang$^{\dagger}$, Chenlu Xiang$^{\dagger}$, Shan Cao$^{\dagger}$, Shugong Xu*$^{\dagger}$, and Jiang Zhu$^{\ddag}$ \\
* Corresponding Author \\
$^{\dagger}$ Shanghai Institute for Advanced Communication and Data Science, \\
Key laboratory of Specialty Fiber Optics and Optical Access Networks, \\
Joint International Research Laboratory of Specialty Fiber Optics and Advanced Communication, \\
Shanghai University, Shanghai, 200444, China\\
$^{\ddag}$ Huawei Technologies, Co. Ltd, Shanghai, 201206, China\\
Email:\{shunqing, xcl, cshan, shugong\}@shu.edu.cn, zhujiang@huawei.com}

\begin{document}
\maketitle

\begin{abstract}
The rapid increasing demand of wireless transmission has incurred mobile broadband to continuously evolve through multiple frequency bands, massive antennas and other multi-stream processing schemes. Together with the improved data transmission rate, the power consumption for multi-carrier transmission and processing is proportionally increasing, which contradicts with the energy efficiency requirements of 5G wireless systems. To meet this challenge, multi-carrier power amplifier (MCPA) technology, e.g., to support multiple carriers through a single power amplifier, is widely deployed in practical. With massive carriers required for 5G communication and limited number of carriers supported per MCPA, a natural question to ask is how to map those carriers into multiple MCPAs and whether we shall dynamically adjust this mapping relation. In this paper, we have theoretically formulated the dynamic carrier and MCPA mapping problem to jointly optimize the traditional separated baseband and radio frequency processing. On top of that, we have also proposed a low complexity algorithm that can achieve most of the power saving with affordable computational time, if compared with the optimal exhaustive search based algorithm.
\end{abstract}

\begin{keywords}
energy efficiency, massive carriers processing, multi-carrier power amplifier, 5G communication
\end{keywords}

\section{Introduction} \label{sect:intro}
In modern wireless communication systems, one of the typical requirements is to deliver the mobile broadband (MBB) \cite{Astely09} transmission in the wireless environments, through either licensed cellular bands or unlicensed bands. With the ever-increasing demand of data communication requirements, enhanced MBB (eMBB) \cite{Liu17}, an evolution of the traditional MBB transmission to support per-user gigabits-per-second transmission, has been identified as one of the major application scenarios in the current developing 5G wireless systems. In order to support this challenging eMBB application, massive multiple-input-multiple-output antenna systems \cite{Larsson14}, millimeter wave communication \cite{Pi16}, and other multi-stream processing schemes such as enhanced carrier aggregation \cite{Lien17} have been proposed. However, the above solutions improves the target data transmission rate at the expense of significant energy consumption, e.g. the power consumption for multi-carrier transmission and processing is proportional to the target data rate, which contradicts with the energy efficiency (EE) requirements of 5G wireless systems.

To solve the energy consumption issues and reduce the potential capital expenditure, multi-carrier power amplifier (MCPA) \cite{Fehri14,Zenteno17,Fujibayashi17}, i.e. to process multiple carriers within a single PA unit, has been proposed as a promising technology for MBB/eMBB communication. The corresponding advantages using MCPA technology are straight-forward. First, by integrating multiple carriers together, the total transmit power becomes greater, which makes MCPAs to operate at higher efficiency areas. Second, the static power consumption of MCPA increases sub-linearly with respect to the number of carriers. Last but not least, from the operator's viewpoint, the deployment and maintenance tasks will be much easier as the number of radio frequency (RF) chains and power amplifiers is greatly reduced. Nevertheless, in the practical systems, the number of carriers supported by MCPA is in general limited and a giant PA to support all the potential carriers is still infeasible based on the current technology.

With massive carriers required for 5G communication and limited number of carriers supported per MCPA, a natural question to ask is how to map those carriers into multiple MCPAs and whether we shall dynamically adjust this mapping relation. Inspired by the results in \cite{Zhang131}, we would like to theoretically answer the above questions through the optimization theory, and the main contributions of this paper are summarized as follows.

\begin{itemize}
\item{\em Theoretical Formulation for Dynamic Mapping} In this paper, based on the practical power amplifier model, we shall formulate the dynamic carrier and MCPA mapping problem as a general optimization problem. Through the theoretical formulation, we establish the relations between the non-linear power efficiency, and the dynamic carrier and MCPA mapping relation, to jointly optimize the traditional separated baseband and RF processing.
\item{\em Low Complexity Dynamic Mapping Algorithm} Due to the non-linear power efficiency and limited carrier and MCPA mapping relations as shown in \cite{Zhang17}, the dynamic mapping problem is in general difficult to solve. In order to enlarge the EE gain and fully utilize the current energy saving technology of MCPA, such as slot-level turn-off, we propose a low complexity algorithm that can dynamically adjust carrier and MCPA mapping relations with affordable computational time.
\end{itemize}

The rest of the paper is organized as follows. Section~\ref{sect:pre} provides the preliminary information on the PA power consumption model and the motivation of dynamic carrier and MCPA mapping scheme. We then formulate the dynamic carrier and MCPA mapping scheme as an optimization problem in Section~\ref{sect:prob}, and propose our low-complexity algorithm in Section~\ref{sect:prop}. The power saving gains of different output power profiles using different mapping algorithms are verified in Section~\ref{sect:sim} and final remarks are given in Section~\ref{sect:conc}.

\section{Background} \label{sect:pre}
In this section, we briefly describe the power model of MCPA used in this paper and a motivating example of dynamic carrier and MCPA mapping, followed by some analytical assumptions.

Consider a general MCPA power model given by $p_{in} = f(p_{out})$, where $p_{in}$ and $p_{out}$ denote the input and output powers of MCPA respectively, and $f(\cdot)$ represents the power efficiency model of MCPA. In the current literature, such as \cite{Xu13}, $f(\cdot)$ is usually modeled as
\begin{eqnarray}
p_{in} = f(p_{out}) = \left\{
\begin{array}{l l}
P_{sta} + \alpha \cdot p_{out}, & (p_{out} > 0) \\
P_{slp}, & (p_{out} = 0)
\end{array} \right.
\end{eqnarray}
where $P_{sta}$ denotes the static power consumption when the power amplifier is active and $P_{slp}$ denotes the power consumption when it is in the sleep mode. $\alpha$ is the linear co-efficient that characterize the dynamic property of power amplifier.

The above power model is valid for conventional Class-A/B/AB power amplifiers and the maximum output power is usually limited. For MCPAs with advanced Doherty technology \cite{Kang17}, the power efficiency in high output power region improves linearly in dB scale and the associated power model needs to be updated as follows,
\begin{eqnarray}\label{eqn:pow_mod}
p_{in} & = & f(p_{out}) \nonumber \\
& = & \left\{
\begin{array}{l l}
\frac{p_{out}}{\beta \cdot 10 \log_{10} (p_{out}) + \gamma}, & ( P_{th} < p_{out} \leq P_{\max})\\
P_{sta} + \alpha \cdot p_{out}, & ( 0 < p_{out} \leq P_{th}) \\
P_{slp}, & (p_{out} = 0)
\end{array} \right.
\end{eqnarray}
where $\beta$ is the linear coefficient that describes the relation between the power efficiency and the output power (in dB scale) and $\gamma$ is the biasing coefficient. $P_{\max}$ and $P_{th}$ denote the maximum output power and the threshold power of MCPAs, respectively.

As a motivating example, we consider the case when four carriers are mapping into two MCPAs and the radiated power of four carriers are 20w, 0w, 20w, and 0w, respectively. Based on the MCPA power model given by \eqref{eqn:pow_mod}, we can show that different mapping schemes\footnote{In the practical systems, typical values for $\alpha$, $\beta$, and $\gamma$, are given by $2.7$, $0.03$, and $-0.06$. $P_{th}$, $P_{\max}$, $P_{sta}$, and $P_{slp}$ are chosen to be 5w, 40w, 20w, and 13w, respectively. Using this model, the mapping relation $(20,0), (20,0)$ requires $2*20/(0.03*10 \log_{10} 20 - 0.06) \approx 2*60.5 = 121$ wats. Similarly, we can compute the power consumption for other mapping relation $(20,20), (0,0)$, which gives 108w.} require 121w and 108w, accordingly. Through this approach, we can save as much as 11\%\footnote{A commercial power saving feature only requires 2\% to 3\% power saving gain in the current market.} of the total required power. It is worth noting that the traditional energy efficient design improves EE at the expense of spectrum efficiency, while the proposed scheme only adjusts the carrier and MCPA mapping relation and achieves power saving gain without reducing the output power as well as the system throughput.

The following assumptions are adopted through the rest of this paper. First, all MCPAs share the same power model as defined in the above. Second, the carrier and MCPA mapping relation can be updated dynamically on a slot basis, and has no effect on the current slot based MCPA sleep mode. Last but not least, we did not consider the power allocation for different carriers in this paper and leave it for further study.

\section{Problem Formulation} \label{sect:prob}
\begin{figure}
\centering
\includegraphics[width = 3.5 in]{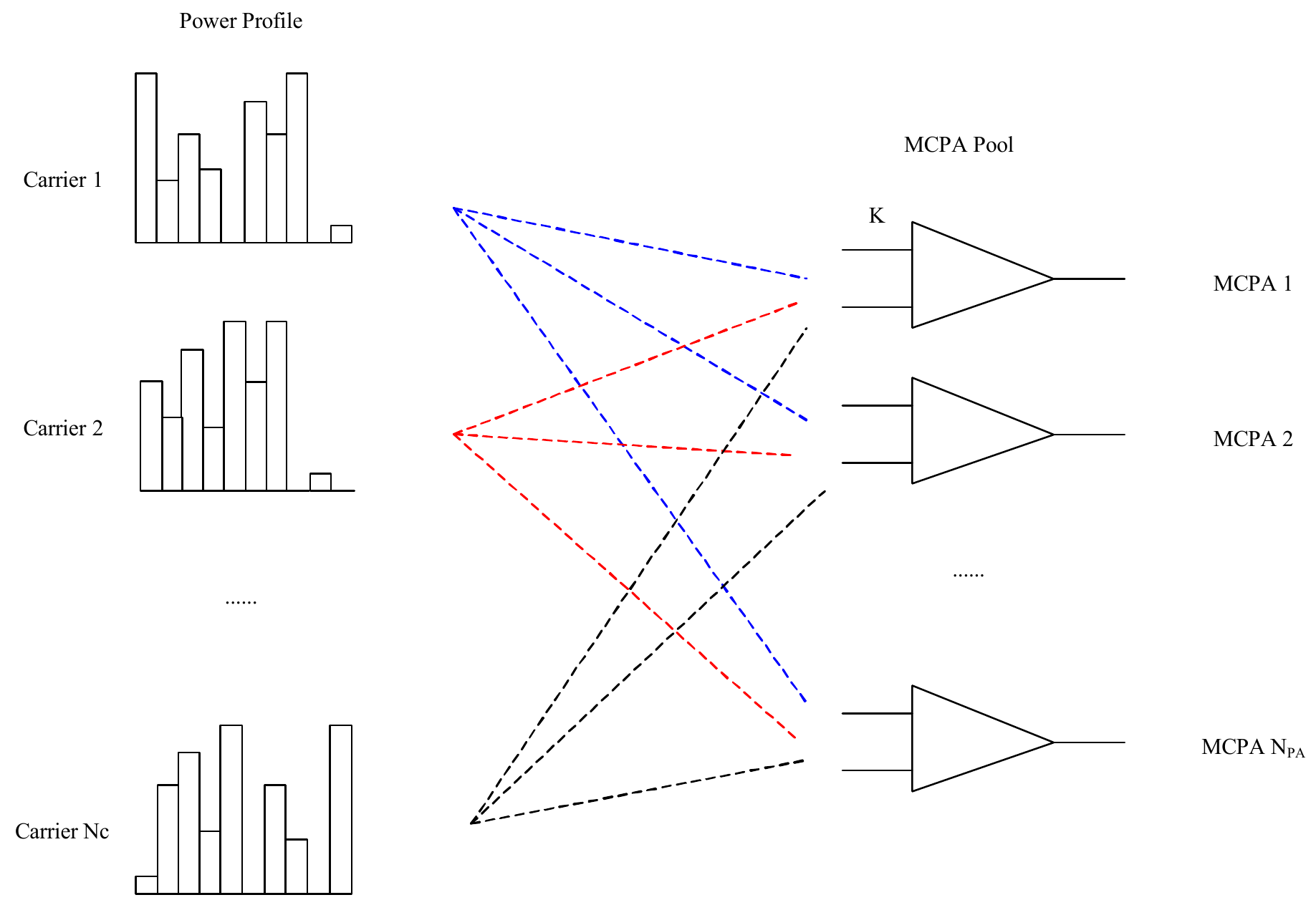}
\caption{Illustration for the dynamic carrier and MCPA mapping scheme, where $N_{C}$ carriers are connecting to $N_{PA}$ MCPAs. Each MCPA supports $K$ carriers at most and based on different transmit power profiles of different carriers, we shall dynamically map those carriers into different MCPAs in order to find the best matching relation.} \label{fig:sys_mod}
\end{figure}

Consider a dynamic carrier and MCPA mapping problem as shown in Fig.~\ref{fig:sys_mod}, where $N_{PA}$ is the number of MCPAs and $N_{C}$ is the total number of carriers. For a given dynamic mapping slot, the mapping relation between the $i^{th}$ carrier and the $j^{th}$ MCPA can be characterized by the variable $c_{i,j}$, where $c_{i,j}$ equals to $1$ if they are connected or $0$ otherwise. The maximum supported carriers per MCPA is denoted by $K$, i.e., $\sum_{i = 1}^{N_C} c_{i,j} \leq K, \forall j$. Denote $p_{i}$ to be the transmission power of the $i^{th}$ carrier, and the power consumption minimization problem can be formulated as,
\begin{eqnarray}
\emph{Problem 1:} \quad \min_{\{c_{i,j}\}} && \sum_{j = 1}^{N_{PA}} f\left(\sum_{i = 1}^{N_{C}} c_{i,j} p_{i}\right)\\
\textrm{s. t.} && c_{i,j} \in \{0, 1\}, \label{eqn:cons11}\\
&& \sum_{i = 1}^{N_C} c_{i,j} \leq K, \quad \forall j, \label{eqn:cons12} \\
&& \sum_{j = 1}^{N_{PA}} c_{i,j} = 1, \quad \forall i. \label{eqn:cons13}
\end{eqnarray}
where the last constraint \eqref{eqn:cons13} guarantees each carrier will be served by one MCPA. Although {\em Problem 1} looks straight-forward, it is never easy to solve due to the following reasons. First of all, the objective function is a non-convex function since the inner function $f(\cdot)$ is non-continuous and non-differentiable. Second, we only have discrete choices of carrier and MCPA mapping relation, which involves complicated exhaustive search to find the global optimal. Finally, the optimization problem needs to be done on a slot basis, each of 0.5ms, where the traditional exhaustive search method is complexity prohibited.

To solve {\em Problem 1}, we first partition those carriers into two parts according to the transmit power $p_i$, denoted by $\mathcal{S}_{ac}$ and $\overline{\mathcal{S}_{ac}}$, where $i \in \mathcal{S}_{ac}$ if $p_i$ is greater than 0 and $i \in \overline{\mathcal{S}_{ac}}$ otherwise. Similarly, MCPAs can be classified as active state part, $\mathcal{S}_{as}$, and sleep state part, $\overline{\mathcal{S}_{as}}$, where the j-th MCPA is in active state, i.e., $j \in \mathcal{S}_{as}$, if the output power $\sum_{i = 1}^{N_{C}} c_{i,j} p_{i}$ is greater than 0, and in sleep state, i.e., $j \in \overline{\mathcal{S}_{as}}$, otherwise. As a result of the power model \eqref{eqn:pow_mod}, we can have an extra benefit in power saving if the corresponding MCPA is in sleep status. Therefore, we can reformulate {\em Problem 1} as follows to fully utilize this property.
\begin{eqnarray}
\min_{\{c_{i,j}\}} && \sum_{j \in \mathcal{S}_{as}} f\left(\sum_{i = 1, i \in \mathcal{S}_c}^{N_{C}} c_{i,j} p_{i}\right) + |\overline{\mathcal{S}_{as}}| \cdot f(0) \\
\textrm{s. t.} && \eqref{eqn:cons11}-\eqref{eqn:cons13} \nonumber
\end{eqnarray}
where $|\cdot|$ denotes the cardinality of the inner set and $f(0) = P_{slp}$ according to \eqref{eqn:pow_mod}. Denote $N_{as}$ and $N_{ac}$ to be the cardinality of $\mathcal{S}_{as}$ and $\mathcal{S}_{ac}$ respectively, and by re-ordering the carriers' and MCPAs' indexes, {\em Problem 1} can be transformed into the following equivalent problem.
\begin{eqnarray}
\min_{\{c_{i',j'}\}} && \sum_{j' = 1}^{N_{as}} f\left(\sum_{i' = 1}^{N_{ac}} c_{i',j'} p_{i'}\right) + (N_{PA} - N_{as}) P_{slp} \label{eqn:prob2}\\
\textrm{s. t.} && c_{i',j'} \in \{0, 1\}, \label{eqn:cons21}\\
&& \sum_{i' = 1}^{N_{ac}} c_{i',j'} \leq K, \quad \forall j', \label{eqn:cons22} \\
&& \sum_{j' = 1}^{N_{as}} c_{i',j'} = 1, \quad \forall i'. \label{eqn:cons23}
\end{eqnarray}
In the above formulation, we have successfully handled the non-continuity of the objective function $f(\cdot)$. However, the remaining problem is still non-convex and difficult to solve.

\section{Proposed Solutions} \label{sect:prop}
In this section, we focus on solving the optimization problem as defined in \eqref{eqn:prob2}-\eqref{eqn:cons23}. By applying the convex relaxation technique and getting rid of the constant part, i.e., $(N_{PA} - N_{as}) P_{slp}$, we can rewrite the above optimization problem as,
\begin{eqnarray}
\emph{Problem 2:} \quad \min_{\{c_{i',j'}\}} && \sum_{j' = 1}^{N_{as}} f\left(\sum_{i' = 1}^{N_{ac}} c_{i',j'} p_{i'}\right) \label{eqn:prob3} \\
\textrm{s. t.} && 0 \leq c_{i',j'} \leq 1, \label{eqn:cons31}\\
&& \eqref{eqn:cons22}, \ \eqref{eqn:cons23}, \nonumber
\end{eqnarray}
where the transmit power $p_{i'}$ is greater than zero after the aforementioned partition process.

To make {\em Problem 2} tractable, we use Taylor expansion to approximate the objective function $f\left(\sum_{i' = 1}^{N_{ac}} c_{i',j'} p_{i'}\right)$ at the middle point $P_{\textrm{mid}} = \frac{P_{\max} - P_{th}}{2}$ and get,
\begin{eqnarray}
& f\left(\sum_{i' = 1}^{N_{ac}} c_{i',j'} p_{i'}\right) \approx f\left(P_{\textrm{mid}}\right) + \frac{f'(P_{\textrm{mid}})}{1!} \Big(\sum_{i' = 1}^{N_{ac}} c_{i',j'} p_{i'} \nonumber \\
& - P_{\textrm{mid}}\Big) + \frac{f''\left(P_{\textrm{mid}}\right)}{2!} \left(\sum_{i' = 1}^{N_{ac}} c_{i',j'} p_{i'} - P_{\textrm{mid}}\right)^2, \label{eqn:prob3_mod}
\end{eqnarray}
where $f'(\cdot)$ and $f''(\cdot)$ denote the first and second order derivative of the function $f(\cdot)$, and we ignore high order terms $\sum_{n=3}^{+\infty}\frac{f^{(n)}\left(P_{\textrm{mid}}\right)}{n!}(\sum_{i' = 1}^{N_{ac}} c_{i',j'} p_{i'} - P_{\textrm{mid}})^n$ since the co-efficient $\frac{f^{(n)}\left(P_{\textrm{mid}}\right)}{n!}$ will quickly tends to be zero when $n$ goes to infinity. Substitute \eqref{eqn:prob3_mod} into \eqref{eqn:prob3}, we can have the following convex optimization problem, which can be solved efficiently through the interior point method\cite{Boyd03}.
\begin{eqnarray}
\emph{Problem 3:} \min_{\{c_{i',j'}\}} N_{as} f\left(P_{\textrm{mid}}\right) + \sum_{j' = 1}^{N_{as}} f'(P_{\textrm{mid}}) \Big(\sum_{i' = 1}^{N_{ac}} c_{i',j'} p_{i'} \nonumber \\
- P_{\textrm{mid}}\Big) + \frac{f''\left(P_{\textrm{mid}}\right)}{2} \left(\sum_{i' = 1}^{N_{ac}} c_{i',j'} p_{i'} - P_{\textrm{mid}}\right)^2 \nonumber \\
\textrm{s. t.} \quad \eqref{eqn:cons22}, \ \eqref{eqn:cons23}, \ \eqref{eqn:cons31} \qquad \qquad \qquad \qquad \qquad \nonumber
\end{eqnarray}
\begin{Rem}
The optimal solution of {\em Problem 3} is only a sub-optimal solution of the original {\em Problem 2} due to the following two reasons. First, we have ignored the high order part as illustrated before, which causes the inaccuracy. Second, we extended the above approximation to the output power region between 0 and $P_{th}$, in which the power model is linear according to \eqref{eqn:pow_mod}.
\end{Rem}

By solving {\em Problem 3}, we can have a real-valued mapping relation $\{c^{*}_{i',j'}\}$ to {\em Problem 2}. Instead of the traditional round operation to project $c^{*}_{i',j'}$ into the discrete set $\{0,1\}$  which may violate constraints \eqref{eqn:cons22} and \eqref{eqn:cons23}, we also proposed a sorting based solution to obtain the result. To be more specific, for each round, we find the maximum value in the solution set $\{c^{*}_{i',j'}\}$. Depending on whether the target MCPA is fully occupied or not, we can map the real-valued solution set into the discrete-valued solution set. After all the above processes, we can obtain the solution for the optimization problem defined by \eqref{eqn:prob2}-\eqref{eqn:cons23}. Note that this problem is equivalent to {\em Problem 1} in the original formulation except for some non-active carriers. Since non-active carriers will not affect the total power consumption as the transmit power per carrier is zero, we can have very flexible mapping relations to unoccupied MCPAs. As a result, we summarize the main procedures of the proposed low complexity dynamic carrier and MCPA mapping scheme in Algorithm~\ref{alg:low}.

\begin{algorithm}[ht]
\caption{Low Complexity Dynamic Carrier and MCPA Mapping Algorithm}
\begin{algorithmic}[1]
\Require the output power requirements $\{P_i\}$, the MCPA power model $f(\cdot)$, the number of MCPAs $N_{PA}$
\Ensure the optimal mapping relation $\{c^{*}_{i,j}\}$
\State Reformulate {\em Problem 1} to the optimization problem defined by \eqref{eqn:prob2}-\eqref{eqn:cons23}.
\State Get rid of non-active MCPAs and formulate {\em Problem 2} through convex relaxation.
\State Further relax {\em Problem 2} to {\em Problem 3} using Taylor expansion.
\State Find the optimal solution $\{c^{*}_{i',j'}\}$ to {\em Problem 3} by the standard convex optimization method, such as interior point.
\For{$i'_{idx} = 1$ to $N_{ac}$ }
\State Find indexes $(i'_{\max}, j'_{\max})$ of the maximum value in the solution set $\{c^{*}_{i',j'}\}$.
\If{$j'^{th}_{\max}$ MCPA has less than $K$ carriers supported.}
\State Set $c^{*}_{i'_{\max},j'_{\max}} = 1$ and $c^{*}_{i'_{\max},j'} = 0$ for all $j' \neq j'_{\max}$ in the solution set $\{c^{*}_{i',j'}\}$.
\Else
\State Set $c^{*}_{i'_{\max},j'_{\max}} = 0$.
\EndIf
\EndFor
\State Obtain the solution set $\{c^{*}_{i',j'}\}$ to the optimization problem defined by \eqref{eqn:prob2}-\eqref{eqn:cons23}.
\State Map back to the solution set for the original {\em Problem 1} $\{c^{*}_{i,j}\}$.
\end{algorithmic}\label{alg:low}
\end{algorithm}

\section{Simulation Results} \label{sect:sim}
In this section, we provide some numerical results to show the effectiveness of the proposed low complexity dynamic carrier and MCPA mapping algorithm. To be more specific, we compare the proposed algorithm with the following baseline schemes. {\em Baseline 1: Static Mapping}, where carriers and MCPAs have static mapping relation. {\em Baseline 2: Exhaustive Search}, where we exploit all the possible dynamic carrier and MCPA mapping relations and find the optimal one. The numerical values of parameters used in the simulation are summarized in Table~\ref{tab:sim}.

\begin{table}[!hbp]%
\caption{Simulation Parameters for Dynamic Carrier and MCPA Mapping} \label{tab:sim}
\centering
\footnotesize
\begin{tabular}{c|c|c|c}
\hline 
\hline
Parameter & Experiment 1 & Experiment 2 & Experiment 3 \\
\hline
$\alpha$  & 2.7 &2.7 & 2.7\\
\hline
$\beta$   & 0.03 & 0.03 & 0.025\\
\hline
$\gamma$  & -0.06 & -0.06 & 0.01\\
\hline
$P_{th}$  & 5 & 5 & 4\\
\hline
$P_{max}$ & 40 & 60 & 40\\
\hline
$P_{sta}$ & 20 & 20 & 14\\
\hline
$P_{slp}$ & 13 & 13 & 9\\
\hline
\hline
\end{tabular}
\end{table}

\begin{figure}
\centering
\includegraphics[width = 3.5 in]{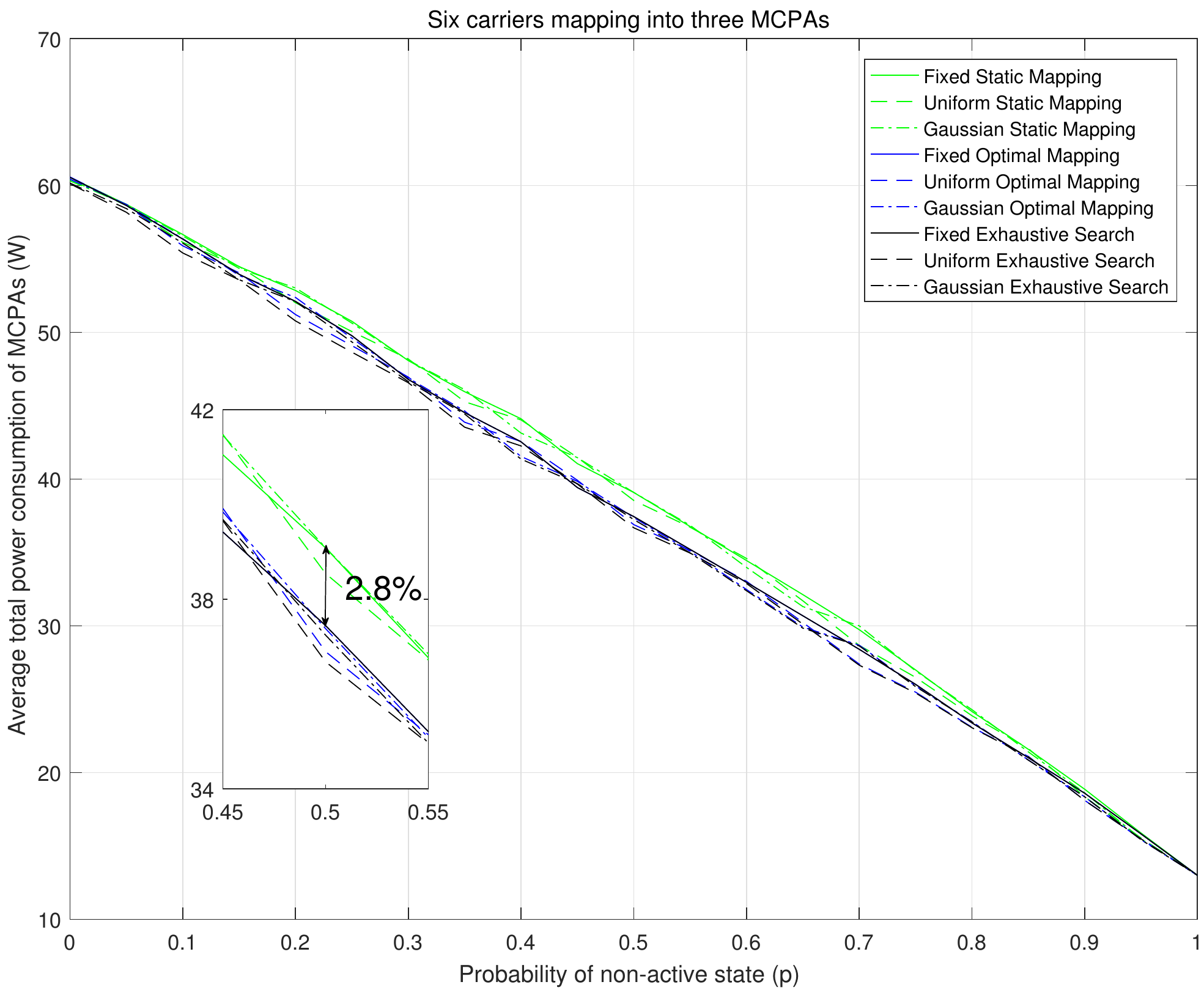}
\caption{Average total power consumption of MCPAs versus probability of non-active carriers. In this experiment, six carriers are mapping into three MCPAs with the maximum supported number $K = 2$. Three types of power profiles are tested with three different carrier and MCPA mapping algorithms. }
\label{fig:test1}
\end{figure}

In the first experiment, we consider a general case where six carriers are mapping into three MCPAs with the maximum supported number $K$ equal to 2. In order to obtain the averaged power saving gain, we run the dynamic carrier and MCPA mapping algorithms over $10^5$ time slots. For each carrier, the probability of non-active state, e.g. $p_{i} = 0$, is assumed to be\footnote{In the network deployment, this parameter is actually used to describe the network loading of cellular systems.} $p$ and in the active state, the output power profile is generated according to certain distributions. In order to have a more complete study, we consider the following three scenarios: 1) fixed transmit power $p_{i} = P_{\max}/2K$, 2) $p_{i}$ generated from uniform distribution between 0 and $P_{\max}/K$, and 3) $p_{i}$ generated from truncated Gaussian distribution with mean $P_{\max}/2K$ and variance $P_{\max}/4K$. The simulated curves for the average power consumption versus the probability of non-active carriers' relation are shown in Fig.~\ref{fig:test1}. By comparing {\em Baseline 1} (green curves) and {\em Baseline 2} (black curves), we can conclude that dynamic carrier and MCPA mapping provides a great potential\footnote{In this case, we only have very limited carrier and MCPA mapping relations, which already satisfied the commercial power saving feature's requirement. Meanwhile, the power saving gain is achieve without any loss in the wireless transmission quality since the output powers for different carriers are {\em NOT} changed. } (2.6\% on average) for power consumption reduction regardless of carrier power profiles, and when the probability of non-active state $p$ is equal to 0.5, the power consumption reduction can reach to as much as 2.8\%. By comparing {\em Baseline 2} (black curves) and the proposed low complexity dynamic carrier and MCPA mapping scheme (blue curves), we can observe that the proposed algorithm actually achieves more than 95\% of power saving gain in most of cases if compared with the optimal carrier and MCPA mapping schemes.

\begin{figure}
\centering
\includegraphics[width = 3.5 in]{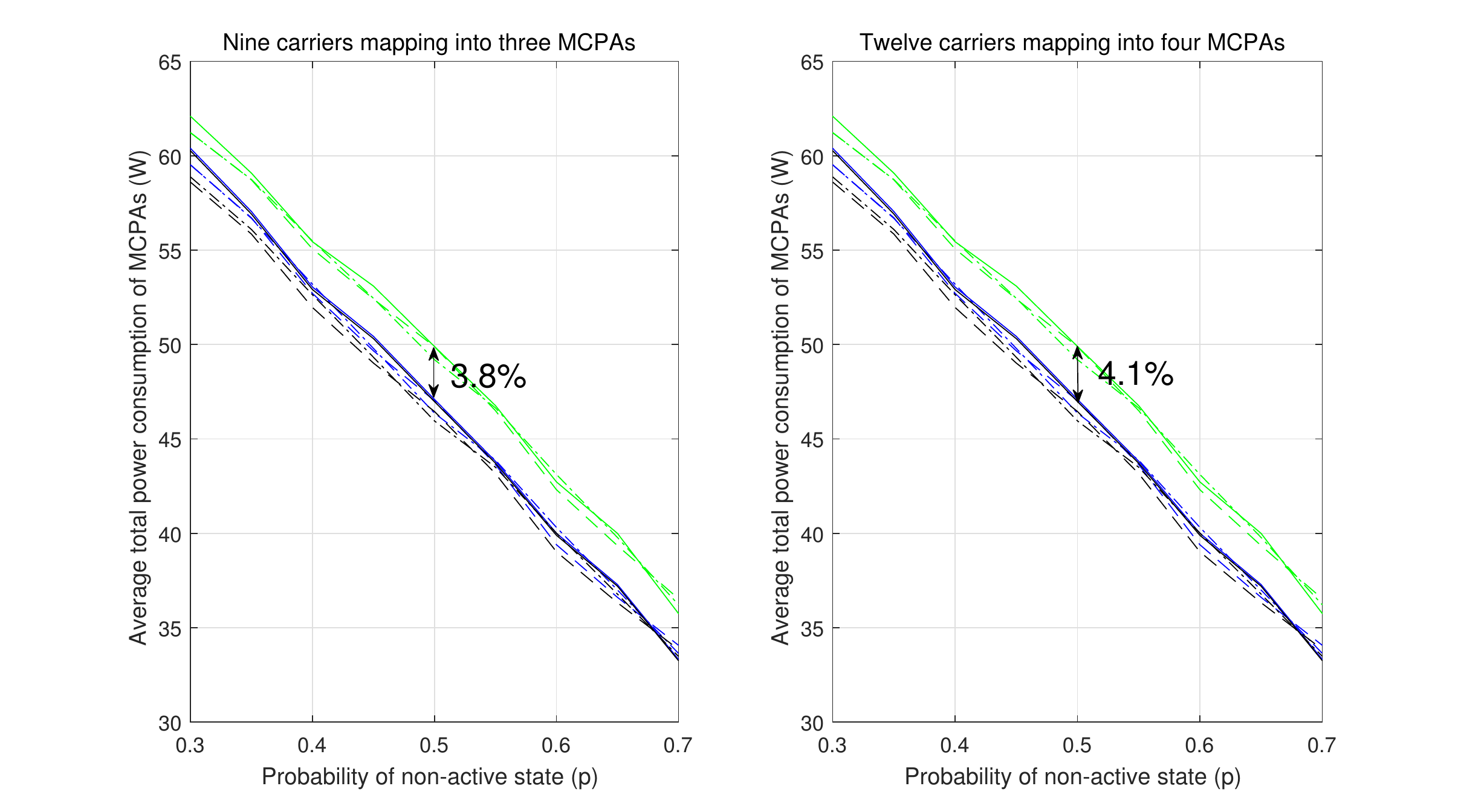}
\caption{Average total power consumption of MCPAs versus probability of non-active carriers. In this experiment, nine / twelve carriers are mapping into three/four MCPAs with the maximum supported number $K = 3$. Three types of power profiles are tested with three different carrier and MCPA mapping algorithms. } \label{fig:test2}
\end{figure}

In the second experiment, we extend the above results to more complicated MCPA with the maximum supported number $K$ equal to 3 and $P_{\max} = 60$w, and the number of MCPAs is assumed to be two and three. As shown in Fig.~\ref{fig:test2}, the power saving gains under the current situation can reach to 3.7\% on average, which is 1.5 times than the previous case ($K = 2$). This is due to the fact that, when the number of supported carriers per MCPA increases, the searching space for different carrier and MCPA mapping relations grows, which provides much more optimization space accordingly. Through this example, we can expect that with more carriers to be supported in 5G wireless systems, dynamic carrier and MCPA mapping scheme will be necessary.

\begin{figure}
\centering
\includegraphics[width = 3.5 in]{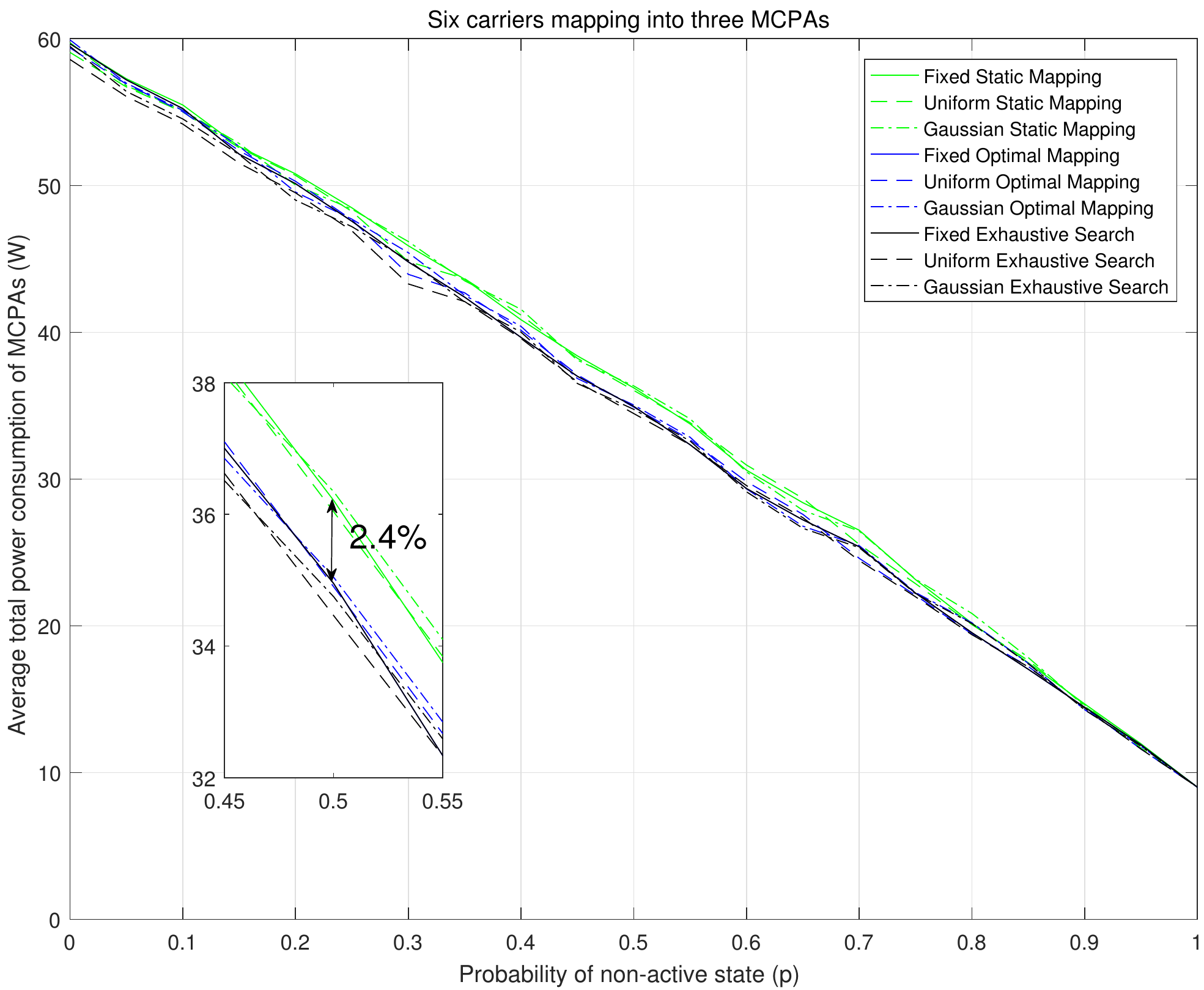}
\caption{Average total power consumption of MCPAs versus probability of non-active carriers. In this experiment, we apply different power models of MCPAs to verify the flexibility of the proposed carrier and MCPA mapping algorithm.} \label{fig:test3}
\end{figure}

In the last experiment, we consider different types of MCPAs, e.g. by slightly tuning the values of parameters used in MCPA power model \eqref{eqn:pow_mod}, to show the flexibility of the proposed low complexity dynamic carrier and MCPAs mapping algorithm. As shown in Fig.~\ref{fig:test3}, although the exact values of parameters are different from the above two experiments, our proposed low complexity dynamic carrier and MCPA mapping algorithm can also provide about 2.5$\%$ power consumption on average and achieve more than 95\% of power saving gain if compared with exhaustive search based algorithms.

\section{Conclusion} \label{sect:conc}
In this paper, we have theoretically formulated the dynamic carrier and MCPA mapping as an optimization problem. Through some convex relaxation techniques, we proposed a low complexity mapping algorithm to dynamically adjust carrier and MCPA relations to reduce the power consumption. From the numerical results, we can show that the proposed algorithm can save as much as 2-5\% percents power consumption reduction and achieve more than 95\% of power saving gain if compared with exhaustive search based algorithms. In summary, we believe the proposed algorithm can play a more important role in future communication systems with massive carriers, and a joint optimization with newly proposed transmission technologies, such as massive multiple-input-multiple-output antenna processing or enhanced carrier aggregation, will be the next step for future research.

\section*{Acknowledgement}
This work was supported by the National Natural Science Foundation of China (NSFC) Grants under No. 61701293, the National Science and Technology Major Project Grants under No. 2018ZX03001009, and research funds from Shanghai Institute for Advanced Communication and Data Science (SICS).

\bibliographystyle{IEEEtran}
\bibliography{IEEEabrv,joint_BB_RF}

\end{document}